\documentclass[lettersize,journal]{IEEEtran}
\usepackage{amsmath,amsfonts}
\usepackage{algorithmic}
\usepackage{array}
\usepackage[caption=false,font=normalsize,labelfont=sf,textfont=sf]{subfig}
\usepackage{textcomp}
\usepackage{stfloats}
\usepackage{url}
\usepackage{verbatim}
\usepackage{graphicx}
\usepackage{arydshln}
\usepackage{booktabs}
\usepackage{color}
\usepackage{hyperref}

\newcommand{\ra}[1]{\renewcommand{\arraystretch}{#1}}

\def\BibTeX{{\rm B\kern-.05em{\sc i\kern-.025em b}\kern-.08em
    T\kern-.1667em\lower.7ex\hbox{E}\kern-.125emX}}
\usepackage{balance}
\begin{document}
\title{Making Flow-Matching-Based Zero-Shot Text-to-Speech Laugh as You Like}

\author{Naoyuki Kanda,
Xiaofei Wang, Sefik Emre Eskimez, Manthan Thakker, Hemin Yang, Zirun Zhu, Min Tang, Canrun Li, Chung-Hsien Tsai, Zhen Xiao, Yufei Xia, Jinzhu Li, Yanqing Liu, Sheng Zhao, Michael Zeng
    \thanks{The authors are with Microsoft Corporation, Redmond, WA 98052, USA.}
    
}


\maketitle

\begin{abstract}
Laughter is one of 
the most expressive and natural aspects of human speech, 
conveying emotions, social cues, and humor. However, most text-to-speech (TTS) systems lack the ability to produce realistic and appropriate laughter sounds, limiting their applications and user experience. 
While there have been prior works to generate natural laughter, 
they fell short in terms of controlling the timing and variety of the laughter to be generated.
In this work, we propose ELaTE, 
a zero-shot TTS that can generate natural laughing speech of any speaker based on a short audio prompt with precise control of laughter timing and expression. Specifically, ELaTE works on the audio prompt to mimic the voice characteristic, the text prompt to indicate the contents of the generated speech, and the input to control the laughter expression, which can be either the start and end times of laughter, or the additional audio prompt that contains laughter to be mimicked. We develop our model based on the foundation of conditional flow-matching-based zero-shot TTS, and fine-tune it with frame-level representation from a laughter detector as additional conditioning. With a simple scheme to mix small-scale laughter-conditioned data with large-scale pre-training data, we demonstrate that a pre-trained zero-shot TTS model can be readily fine-tuned to generate natural laughter with precise controllability, without losing any quality of the pre-trained zero-shot TTS model. 
Through 
objective and subjective 
evaluations, we show that ELaTE can generate laughing speech with significantly higher quality and controllability compared to conventional models. See \url{https://aka.ms/elate/} for demo samples.
\end{abstract}

\begin{IEEEkeywords}
Zero-shot text-to-speech, laughter generation, speech-to-speech translation.
\end{IEEEkeywords}

\section{Introduction}

Human speech contains not only linguistic information but also non-linguistic cues such as laughter, crying, and whispering. These cues are used to express various feelings and intentions in communication. While the text-to-speech (TTS) system has improved significantly \cite{ren2019fastspeech,ren2020fastspeech,kim2020glow,kong2020hifi}, 
most TTS systems still lack the ability to generate such cues. In this work, we focus particularly on laughter generation, one of the most expressive and natural aspects of human speech. It is beneficial to make the TTS system laugh for many applications, such as conversational agent systems that make close communication with the user, or speech-to-speech translation systems that can convey the emotion of the source audio.

Various attempts have been made to generate natural laughter sounds from TTS systems. One direction is to represent the laughter by a specific linguistic unit, such as one of the phonemes~\cite{el2015speech,nagata2018defining} or a special label representing laughter~\cite{xin2023laughter}.
However, such a model
suffers from the lack of controllability 
and the limited expression,
as it generates laughter that is randomly determined by the statistical model.
It is unsuitable for a scenario where we want to control the laughter precisely, such as in a speech-to-speech translation scenario where we aim to accurately transfer the laughter from the source audio.
To enhance controllability, Mori et al.~\cite{mori2019conversational} proposed the utilization of power contours of laughter. Luong et al.~\cite{luong2021laughnet} incorporated the silhouette of laughter to achieve finer control over the generated laughter. Despite these advancements, existing methods still face limitations in precisely controlling various expressions of laughter, such as subtle chuckles, uproarious guffaws, or mischievous snickers. 
Furthermore,
these methods are not applicable to generating speech where the speaker talks while laughing.
Recently, Seamless Expressive~\cite{barrault2023seamless} proposed a speech-to-speech translation system 
that can transfer the expressive attributes, including laughter, in the source audio when it generates translated audio. 
However, it represents the expressiveness feature with a single vector per utterance, 
which results in a lack of control over the timing of laughing.

Along with the progress of laughter generation, there has been recent progress in the zero-shot TTS system that can generate speech of any speaker based on a short audio prompt from the target speaker~\cite{wang2023neural,kharitonov2023speak,le2023voicebox,shen2023naturalspeech,jiang2023mega,jiang2023mega2,wang2023speechx,yang2023uniaudio,liu2023generative}. VALL-E~\cite{wang2023neural} proposed to formulate the zero-shot TTS as the language modeling based on a neural audio codec, and showed impressive zero-shot TTS ability. VoiceBox~\cite{le2023voicebox} proposed to train a conditional flow-matching-based TTS system based on a masked audio infilling task, and showed superior speaker similarity and intelligibility over VALL-E. 
Several extensions of zero-shot TTS models have also been proposed, such as
generative pre-training~\cite{liu2023generative}, multi-task training~\cite{wang2023speechx,yang2023uniaudio}, 
large language model integration~\cite{hao2023boosting}.
Nevertheless, the existing zero-shot TTS systems still lack controllability regarding the speech expression for the generated audio. For example, existing zero-shot TTS systems do not necessarily generate laughing speech even when the audio prompt includes laughter. They also lack the ability to decide when and how to laugh, which is essential for many applications such as speech-to-speech translation.

To address these issues, we need a zero-shot TTS system that can adjust the speech expression of the output audio according to the user’s preference.
In pursuit of this objective,
we propose 
{\bf ELaTE}\footnote{{\bf E}xpressive {\bf La}ughter-controllable Zero-shot {\bf T}ext-to-speech {\bf E}ngine.},
a zero-shot TTS
system 
that can generate natural laughing speech based on three inputs: a speaker prompt to mimic the voice characteristic, a text prompt to indicate the contents of the generated speech, and an additional input to control the laughter expression (Fig~\ref{fig:main}). 
The laughter, including the choice not to laugh, can be controlled either by specifying the start and end times for laughing or by using an additional audio prompt containing laughter to be mimicked. 
We develop our model based on the conditional flow-matching-based zero-shot TTS and fine-tune it with a frame-level laughter indicator as additional conditioning. 
To alleviate the degradation of the quality of the pre-trained zero-shot TTS model,
we introduce a simple scheme to mix the small-scale laughter-conditioned data and large-scale pre-training data in the fine-tuning stage.
To evaluate our model, we curated Chinese-to-English speech-to-speech translation testing samples including laughter from DiariST-AliMeeting data~\cite{yang2023diarist}.
Through our 
objective and subjective 
evaluation using LibriSpeech \cite{panayotov2015librispeech} and DiariST-AliMeeting laughter test set, we show that our model can generate controlled laughter from any speaker
that is significantly better than the prior model, 
without losing any quality of the baseline zero-shot TTS system. 

The key properties of ELaTE, which constitute its novelty to the prior zero-shot TTS models, are as follows. 
\begin{itemize}
\item {\bf Precise control of laughter timing:} A user can specify the timing for laughter, which critically affects the nuance of the generated speech. 
ELaTE can generate a speech where the speaker laughs while talking when instructed to do so.
\item {\bf Precise control of laughter expression:} A user can guide the laughter expression using an example audio containing laughter. This feature is especially useful for speech-to-speech translation, where mimicking the laughter in the source audio is vital for accurately conveying its nuance.
\item {\bf Build upon a well-trained zero-shot TTS:} 
ELaTE can generate natural speech without compromising audio quality and with a negligible increase in computational cost compared to the conventional zero-shot TTS model. 
When the laughter prompt is not provided, ELaTE works the same as the conventional zero-shot TTS.
\end{itemize}
We encourage the reader to listen to our samples on the demo page \url{https://aka.ms/elate/}.

\section{Related Work}

\subsection{Zero-shot TTS}

Zero-shot TTS is a technique that enables the generation of an arbitrary voice with minimal enrolled recordings, or an audio prompt, without the need for re-training model parameters. This technique has a wide range of applications, such as speech-to-speech translation, personal assistant services, news broadcasting, audio navigation, etc.

There has been a surge of research in this area. Early studies leveraged speaker embeddings as additional conditioning to the TTS system~\cite{arik2018neural,jia2018transfer}. More recently, VALL-E~\cite{wang2023neural} proposed formulating the zero-shot TTS problem as a language modeling problem in the neural codec domain. They demonstrated that zero-shot TTS could be achieved by leveraging the in-context learning capabilities of robust language models.
NaturalSpeech2~\cite{shen2023naturalspeech} proposed a method to estimate the latent vector of a neural audio codec using the in-context learning capability of diffusion models.
Voicebox~\cite{le2023voicebox}, a non-autoregressive flow-matching model, is trained to infill speech given audio context and text. It can be utilized for mono or cross-lingual zero-shot text-to-speech synthesis, noise removal, content editing, style conversion, and diverse sample generation.
Several extensions and improvements have been proposed, such as disentangled representation learning~\cite{jiang2023mega,jiang2023mega2}, generative pre-training~\cite{liu2023generative}, multi-task training~\cite{wang2023speechx,yang2023uniaudio}, and integration with large language models~\cite{hao2023boosting}.

Our work builds upon the foundation laid by Voicebox, where we employ a flow-matching-based audio model for speech generation. We demonstrate that a pre-trained flow-matching-based zero-shot TTS model can be effectively fine-tuned to generate natural laughter with precise control over timing and laughter expression.

\begin{figure}[t]
  \centering
  \includegraphics[width=1.0\linewidth]{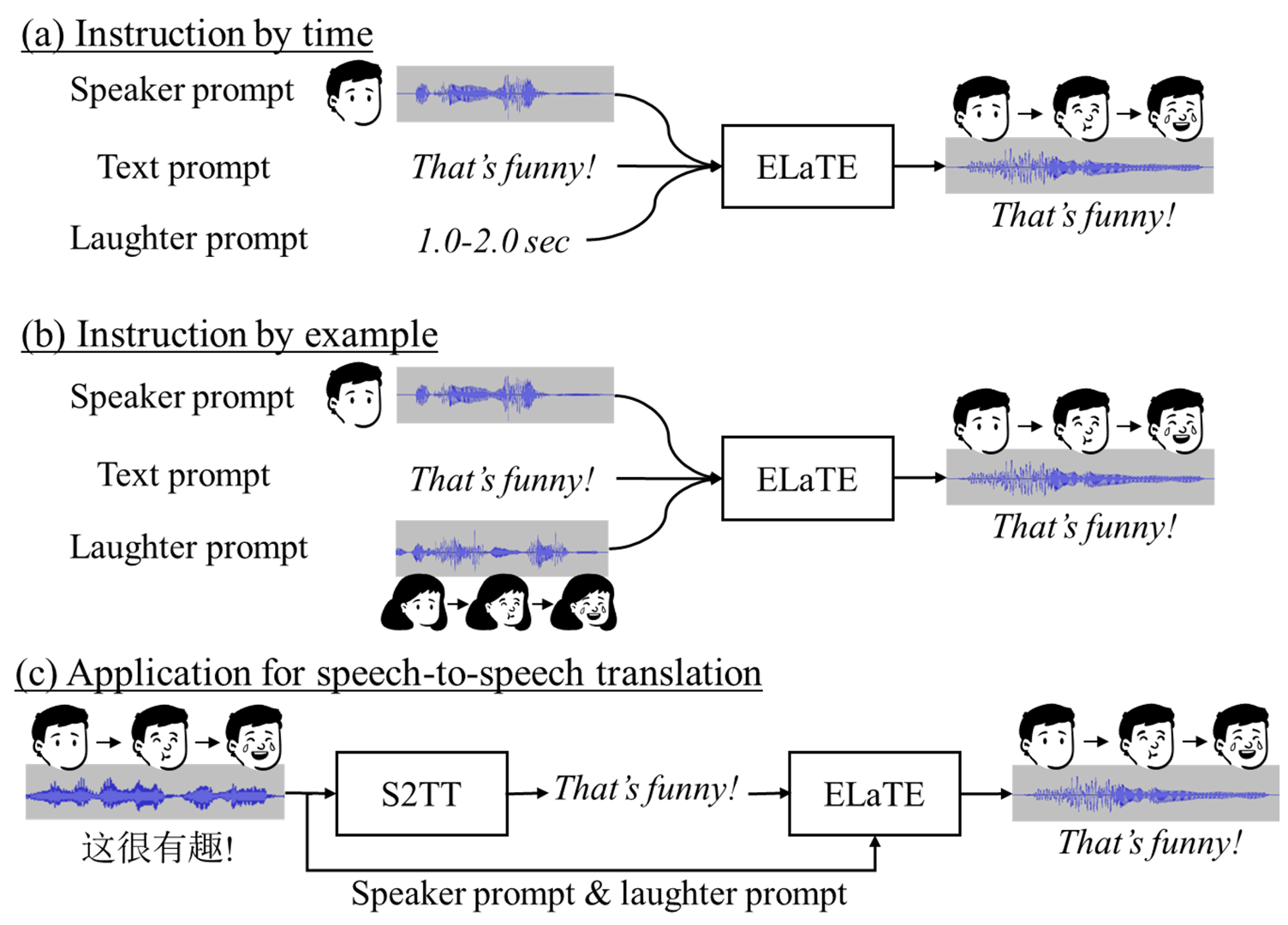}
      \caption{An overview of the capability of ELaTE. ELaTE can generate natural laughing speech
      from a speaker prompt to mimic the voice characteristic,
      a text prompt to indicate the contents of the generated speech,
     and an additional input to control the laughter expression. 
     The laughter, including the choice not to laugh, can
be controlled either by (a) specifying the start and end times
for laughing or (b) by using an audio example that
contains laughter to be mimicked.
(c) ELaTE is particularly beneficial for speech-to-speech translation that precisely transfers the nuance in the source speech. This is achieved by combining it with speech-to-text translation (S2TT).
}
  \label{fig:main}
\end{figure}

\subsection{Conditioning generative models}

Conditioning generative models is a common practice across various research fields. For example, TTS itself is a speech generation technique that is conditioned by a text input. It is also common to condition the TTS model further by adding prosody information such as pitch and energy~\cite{ren2020fastspeech,huang2023holistic}.

Recently, researchers have been exploring ways to inject additional control into well-trained generative models. For instance, in image generation, models are often conditioned on signals such as the human pose skeleton or the edge map to guide the generation process~\cite{zhang2023adding,mou2023t2i}. This approach enables precise control over the generated content, facilitating the creation of images that closely align with the desired output. A similar approach has been applied to music generation, where authors inject elements like melody, pitch, and dynamics into the music generation model~\cite{wu2023music}. Our work extends these concepts to the domain of zero-shot TTS. 
To the best of our knowledge, our work is the first to inject additional conditioning of a frame-level expression signal (in our case, frame-level laughter expression) into a well-trained zero-shot TTS model to precisely control speech generation.
This method allows us to generate speech that not only matches the voice characteristics of a given speaker but also includes natural-sounding laughter at specified intervals with a specified expression.

\begin{figure*}[t]
  \centering
  \includegraphics[width=1.0\linewidth]{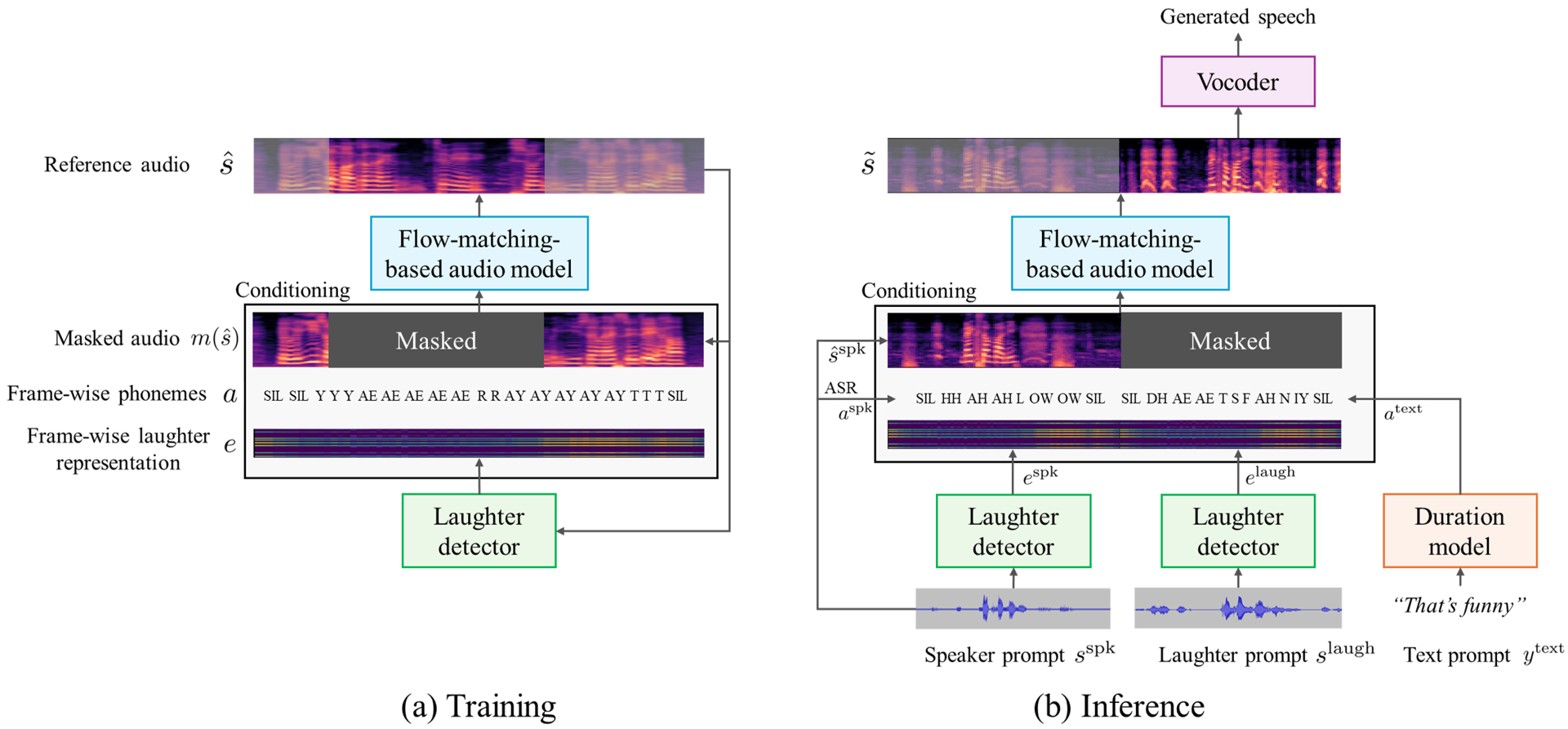}
      \caption{An overview of (a) training and (b) inference of the flow-matching-based zero-shot TTS with laughter expression control.}
  \label{fig:overview}
\end{figure*}

\section{Method}
\subsection{Overview}

Fig.~\ref{fig:overview} illustrates an overview of the training and inference scheme of the proposed zero-shot TTS. Suppose we have a training audio sample $s$
and its transcription $y$. 
We extract the log mel-filterbank feature $\hat{s}\in\mathbb{R}^{F\times T}$ from $s$,
where $F$ is the feature dimension, and $T$ is the length of the feature sequence.
We can also generate a frame-wise phoneme alignment $a\in\mathbb{Z}_{+}^{T}$ 
using a conventional forced alignment algorithm. In addition to these, we assume we have a laughter detector model, from which we can generate a frame-wise laughter representation $e\in\mathbb{R}^{D^{\rm laugh}\times T}$, 
either as laughter probabilities ($D^{\rm laugh}=1$) 
or laughter embeddings ($D^{\rm laugh}\in\mathbb{N}$) from the intermediate state of the laughter detector model. We further assume a masking function $m(\cdot)$, which randomly zeroes out a part of $\hat{s}$. 
The model training follows the speech infilling task proposed in~\cite{le2023voicebox}, where we train a conditional flow-matching model to estimate the conditional distribution $P(\hat{s}|m(\hat{s}),a,e)$.

During inference, we feed a model three inputs, namely, a speaker prompt $s^{\rm spk}$ to mimic the voice characteristic, a text prompt $y^{\rm text}$ to indicate the contents of the generated speech, and a laughter prompt $s^{\rm laugh}$ to control the laughter expression. 
The laughter detector model is applied to both $s^{\rm spk}$ and $s^{\rm laugh}$ to extract 
the corresponding laughter representation $e^{\rm spk}\in \mathbb{R}^{D^{\rm laugh}\times T^{\rm spk}}$ and $e^{\rm laugh}\in \mathbb{R}^{D^{\rm laugh}\times T^{\rm laugh}}$, respectively. 
The automatic speech recognition (ASR) is applied to $s^{\rm spk}$ 
to obtain a frame-wise phoneme sequence $a^{\rm spk}\in \mathbb{Z}_{+}^{T^{\rm spk}}$. 
In addition, a duration model will predict a frame-wise phoneme sequence 
$a^{\rm text}\in \mathbb{R}^{T^{\rm text}}$ for the text prompt $y^{\rm text}$.
Here,
$a^{\rm text}$ and $e^{\rm laugh}$ are configured to have 
the same length (i.e. 
$T^{\rm text} = T^{\rm laugh}$) by either linearly scaling 
the length of $a^{\rm text}$ when 
the estimated duration is longer than the duration of the laughter prompt
or filling silence phoneme to $a^{\rm text}$ when 
the estimated duration is shorter than the duration of the laughter prompt.
Finally, the log mel-filterbank feature $\hat{s}^{\rm spk}\in \mathbb{R}^{F\times T^{\rm spk}}$ is computed 
from $s^{\rm spk}$.
The flow-matching-based audio model will generate 
mel-filterbank features $\tilde{s}$ based 
on the learnt distribution of $P(\tilde{s}|[\hat{s}^{\rm spk}; z^{\rm text}], [a^{\rm spk}; a^{\rm text}], [e^{\rm spk}; e^{\rm laugh}])$, 
where $z^{\rm text}$ is an all-zero matrix with a shape of ${F\times T^{\rm text}}$,
and $[;]$ is a concatenation operation in the dimension of $T^*$.
The generated features $\tilde{s}$ are then converted to
the speech signal based on the vocoder.

\subsection{Conditional flow-matching}

We use conditional flow-matching~\cite{lipman2022flow} as the backbone of our audio model. This model is a type of generative model that uses continuous normalizing flows~\cite{chen2018neural} 
to transform a simple prior distribution $p_0$ into a complex one $p_1$ that fits the data.
Specifically, given a data point $x$, 
a neural network with parameter $\theta$ models
a time-dependent vector field $v_t(x;\theta)$,
which is 
used to construct 
 a flow $\phi_t$, which is then used to reshape
the prior distribution to the target distribution.
Lipman et al.~\cite{lipman2022flow} proposed to train such a neural network 
with the conditional flow-matching objective,
\begin{equation}
\mathcal{L}^{\rm CFM}(\theta)=\mathbb{E}_{t,q(x_1), p_t(x|x_1)}||u_t(x|x_1)-v_t(x;\theta)||^2,
\end{equation}
where
$x_1$ is the random variable representing the training data,
$q$ is training data distribution,
$p_t$ is a probability path at the time step $t$,
and $u_t$ is the corresponding vector field to $p_t$.
Lipman et al. also suggested a conditional flow 
called optimal transport path
with a form of 
$p_t(x|x_1)=\mathcal{N}(x|tx_1, (1-(1-\sigma_{\rm min})t)^2I)$
and
$u_t(x|x_1)=(x_1-(1-\sigma_{\rm min})x)/(1-(1-\sigma_{\rm min})t)$.
Given the promising result presented in~\cite{le2023voicebox},
we build our audio model based on these foundations.

\subsection{Model architecture}

{\bf Flow-matching-based audio model:} We use Transformer with U-Net style skip connection~\cite{le2023voicebox}
as a backbone of the flow-matching-based audio model. 
The input to the audio model is $\hat{x}$, $a$, $e$,
the flow step $t$,
and noisy speech $x_t$.
The phoneme alignment $a$ is first converted to phoneme embedding 
sequence $\hat{a}\in\mathbb{R}^{D^{\rm phn}\times T}$.
Then, $\hat{x}$, $x_t$, $\hat{a}$, $e$ are all stacked to form 
a tensor with a shape of $(2\cdot F+D^{\rm phn}+D^{\rm laugh})\times T$,
followed by a linear layer to output a tensor with a shape of $F\times T$.
Finally, an embedding representation, $\hat{t}\in\mathbb{R}^F$, of $t$ is appended
to form the input tensor with a shape of $\mathbb{R}^{F\times (T+1)}$ to the Transformer.
The Transformer is trained to output a vector field $v_t$ with the conditional flow-matching objective $\mathcal{L}^{\rm CFM}$.

{\bf Duration model:} The regression-based duration model we utilize aligns closely with the one presented in~\cite{le2023voicebox}. Given a phoneme sequence $a$, we convert it into a duration sequence $l$, where each element represents the duration of the corresponding phoneme in $a$. A Transformer model is trained to estimate $l$ from its masked version and unmasked phoneme embeddings. The training of the duration model is based on the $L_2$ regression loss over the masked region.

{\bf Laughter detector:} We employ a laughter detector model, as described in~\cite{ryokai2018capturing,gillick2021robust}\footnote{\url{https://github.com/jrgillick/laughter-detection}}, to extract laughter representations. This model, built on ResNet~\cite{he2016deep}, is trained on the Switchboard corpus to generate laughter probabilities at a frame rate of 43.1 fps. We extract laughter embeddings from the model's final layer, yielding 32-dimensional embeddings ($D^{\rm laugh} =32$). To align with the audio model's frame rate, we apply a linear scaling to the frame rate.

\subsection{Fine-tuning pre-trained zero-shot TTS with additional conditioning}

In this work, we fine-tune a well-trained zero-shot TTS model using a small training dataset that includes laughter. The base model, which was trained on large-scale data, did not initially include the frame-wise laughter representation 
$e$. During the fine-tuning stage, we accommodate the additional input of 
$e$ by extending the weight and bias of the linear layer to form the Transformer input by the dimension of 
$e$, initializing only the extended parameters randomly.

Introducing additional control into generative models often presents a challenge due to the scarcity of training data. Naively fine-tuning an existing model with a limited amount of conditioning training data can significantly compromise the model’s quality. Both ControlNet~\cite{zhang2023adding} and the T2I adapter~\cite{mou2023t2i} propose fine-tuning an additional auxiliary network while keeping the base model frozen to prevent catastrophic degradation. These methods were devised under the assumption that the original pre-training data is inaccessible, and they come with the cost of a small but noticeable increase in the number of parameters.

In contrast, we employ a more straightforward approach by fine-tuning an existing model with a combination of unconditioned pre-training data and conditioned fine-tuning data, including laughter, assuming that we have access to the original pre-training data. For the pre-training data, 
$e$ was set to be an all-zero tensor. This method provides us with reasonable controllability with almost no additional parameters, and without any compromise on the output quality.
We chose this approach because, in our preliminary experiment, we did not observe promising results from the approach of freezing the base model, which was trained by read speech (in our case, LibriLight~\cite{kahn2020libri}). We speculate that our base model did not have sufficient ability to generate laughter due to the scarcity of laughter speech in the read speech corpus. On the other hand, fine-tuning the base model with a mixture of unconditioned pre-training data and conditioned fine-tuning data can expose the model to laughter, enabling it to learn. Exploring better architectures or optimization techniques for our model is part of our future work.

\section{Experiments}
\label{sec:experiments}

\subsection{Training data}

{\bf Pre-training data:}
The base audio model and the duration model were trained using LibriLight~\cite{kahn2020libri}, which comprises 60 thousand hours of untranscribed English reading speech from over 7,000 speakers. As LibriLight does not provide reference transcriptions, we utilized an off-the-shelf Kaldi model\footnote{\url{https://kaldi-asr.org/models/m13}}, trained on the 960-hour Librispeech data~\cite{panayotov2015librispeech} with 3x speed perturbation~\cite{ko2015audio}, to transcribe the LibriLight training data.

{\bf Fine-tuning data:}
Our fine-tuning data, which contains laughter speech, was curated from the AMI meeting corpus~\cite{carletta2005ami}, Switchboard corpus~\cite{godfrey1992switchboard}, and Fisher corpus~\cite{cieri2004fisher}. We selected all utterances marked with laughter from each corpus's reference transcriptions, resulting in a total of 459.8 hours of speech containing laughter. It should be noted that the fine-tuning data also contains a substantial amount of neutral speech, as people tend to laugh only at certain parts of the speech.

\subsection{Model and training configurations}

Our audio model was built using a transformer architecture with U-Net style skip connections~\cite{le2023voicebox}. It included 24 layers, 16 attention heads, an embedding dimension of 1024, a feed-forward layer dimension of 4096, and a dropout rate of 0.1. The total number of model parameters was 335 million.
The model was based on
a 100-dim log mel-filterbank 
from 24kHz sampling audio\footnote{All training data was upsampled to 24kHz. In the inference stage, we upsampled the speaker prompt to 24kHz, generated the audio, and then downsampled it to 16kHz. These settings were chosen to align with the 24kHz sampling rate of the vocoder we used.}
at every 10.7 (=256/24000) msec,
and a MelGAN-based vocoder~\cite{kumar2019melgan} was used to 
convert the log mel-filterbank into
a speech signal.
In the pre-training phase, 
we followed the masking strategy
proposed in~\cite{le2023voicebox}.
The model was trained for 390K steps with an effective mini-batch size of 307,200. We adopted a linear learning rate schedule with a warm-up phase for the first 20K updates, reaching a peak at $7.5\times10^{-5}$. Additionally, we applied speed perturbation~\cite{ko2015audio} with a ratio of {0.9, 1.0, 1.1} to enhance speaker diversity.
During the fine-tuning phase, the model underwent an additional 40K training steps with the same effective mini-batch size of 307,200 and the same masking strategy. We used a linear decay learning rate schedule with the peak learning rate at $7.5\times10^{-5}$.

For our duration model, we adhered closely to the settings outlined in~\cite{le2023voicebox,vyas2023audiobox}. The model was a Transformer with 8 layers, 8 attention heads, an embedding dimension of 512, a feed-forward layer dimension of 2048, and a dropout rate of 0.1. It was trained for 600K updates with an effective batch size of 120K frames.

\subsection{Evaluation data}

{\bf DiariST-AliMeeting laughter test set:} To evaluate the zero-shot TTS capability in generating laughing speech, we established a new experimental setting based on the speech-to-speech translation scenario. First, we curated 154 Chinese utterances that included laughter\footnote{The list of 154 utterances, including natural laughter, along with their transcription and translation, can be found on our demo page \url{https://aka.ms/elate/}.} from the evaluation subset of the DiariST-AliMeeting test set~\cite{yang2023diarist}\footnote{\url{https://github.com/Mu-Y/DiariST}}. This test set comprises Chinese meeting recordings along with their human transcriptions and English translations.
Unlike the staged dataset by a small number of speakers like Expresso~\cite{nguyen2023expresso}, our test set contains spontaneous laughter from 24 speakers,
where the speaker often laughs during only a small part of the utterance.
In the evaluation process, we first applied a speech-to-text translation model to the Chinese test set to obtain the English translation. Subsequently, the TTS model was utilized, using the English translation as the text prompt and the Chinese audio as the speaker and laughter prompts. The generated speech is expected to be an English-translated speech with the same speaker and laughter characteristics. We computed several 
objective and subjective 
metrics on the generated speech, which are described in the following subsection.

{\bf LibriSpeech test-clean:} To evaluate the zero-shot TTS capability under neutral speech conditions, we also assessed our model using the ``test-clean'' set from LibriSpeech. Following the cross-sentence zero-shot TTS condition~\cite{wang2023neural}, we used audio clips ranging from 4 to 10 seconds in length from the ``test-clean'' set. The speech was then generated using a 3-second clip from another sample of the same speaker.

\begin{table}[t]
\ra{1.0}
  \caption{
  Evaluation of zero-shot TTS systems for LibriSpeech test-clean with cross-utterance setting.  In this experiment, a zero vector was fed to the laughter conditioning to validate the neutral behavior of the model. 
  $^\dagger$ Numbers are taken from~\cite{le2023voicebox}.}
  \label{tab:test-clean}
 \vspace{-3mm}
  \centering
{ \footnotesize
\begin{tabular}{@{}lcc@{}}
    \toprule 
Model & WER$\downarrow$ & Speaker SIM-o$\uparrow$ \\
\midrule
Ground-truth & 2.1 & 0.711 \\ \midrule
A$^3$T~\cite{bai20223}$^\dagger$& 63.3 & 0.046 \\
YourTTS~\cite{casanova2022yourtts}$^\dagger$& 5.9 & 0.337 \\
Voicebox~\cite{le2023voicebox} & {\bf 1.9} & 0.662 \\
\midrule
Our pre-trained TTS & 2.2  & 0.653 \\
\hspace{2mm}$\hookrightarrow$ FT (= Baseline TTS) & 2.2  & 0.658  \\
\hspace{2mm}$\hookrightarrow$ FT with laughter prob. (= ELaTE (prob.)) & 2.2  & {\bf 0.663}  \\
\hspace{2mm}$\hookrightarrow$ FT with laughter emb. (= ELaTE (emb.))& 2.2 & 0.662  \\
                \bottomrule
  \end{tabular}
  }
\end{table}

\subsection{Objective evaluation metrics}
We used objective metrics as described below.

{\bf Word error rate (WER):}  To assess the intelligibility of the generated audio, we employed an ASR model, and computed the WER. In this study, we used a Hubert-large based ASR~\cite{hsu2021hubert}, following prior works~\cite{wang2023neural,le2023voicebox}.

{\bf Speaker SIM-o:} 
Speaker SIM-o computes the cosine similarity between the speaker embeddings from the original speaker prompt and the generated audio. In this work, we employed a WavLM-large-based speaker verification model~\cite{chen2022wavlm}\footnote{\url{https://github.com/microsoft/UniSpeech/tree/main/downstreams/speaker_verification}}, following prior works~\cite{wang2023neural,le2023voicebox}.

{\bf ASR-BLEU:} 
ASR-BLEU is a measure used to assess the overall quality of speech-to-speech translation. The BLEU score~\cite{papineni2002bleu} is computed based on the ASR transcription of the generated speech. We used Whisper-v2~\cite{radford2022robust} for the ASR system and SacreBLEU version 2.3.1~\cite{post2018call} for the BLEU computation.

{\bf AutoPCP:} 
AutoPCP~\cite{barrault2023seamless} is a model-based estimator of PCP~\cite{huang2023holistic}, which measures how similarly two speech samples sound
in prosody. We used the scoring tool provided by 
the authors\footnote{\url{https://github.com/facebookresearch/stopes/tree/main/stopes/eval/auto_pcp}}.

{\bf Laughter timing:} 
To compute the laughter timing similarity in the speech-to-speech translation scenario, we applied the laughter detection model~\cite{ryokai2018capturing,gillick2021robust}, the same one used for our model training, to predict the frame-wise laughter probability for the laughter prompt and the generated audio. We then computed the Pearson correlation coefficient between the two laughter probability sequences.

{\bf Laughter SIM:} 
To assess the similarity of laughter between the laughter prompt and the generated audio, we first extracted frame-wise laughter embeddings from the final layer of the laughter detector model for both audios. We then computed the weighted average of the cosine similarity of the two laughter embeddings, where the weight is derived from the laughter probability from the laughter detector.

\subsection{Subjective evaluation metrics}
\label{sec:subjective_metric}
In addition to these objective metrics, we also conducted a subjective evaluation. We evaluated the systems in the following aspects:

{\bf Naturalness mean opinion score (NMOS)}: The naturalness of the generated speech from 1 (bad) to 5 (excellent) with 1-point increments.

{\bf Speaker similarity MOS (SMOS)}: The similarity between the speaker prompt and the generated speech
from 1 (not at all similar) to 5 (extremely similar) with 1-point increments.
We followed the scoring guidance in~\cite{vyas2023audiobox}.

{\bf Laughter similarity MOS (LMOS)}: The similarity of laughter between the laughter prompt and the generated speech from 1 (not at all similar) to 5 (extremely similar) with 1-point increments. We used the following scoring guidance:
1: Not at all similar: The laughter timing and characteristics are completely different, or laughter is not hearable. 
2: Slightly similar: The laughter timing and characteristics have minimal similarities but are mostly characterized by noticeable differences. 
3: Moderately similar: The laughter timing and characteristics have some shared characteristics and also some noticeable differences, in equal parts. 
4: Very similar: The laughter timing and characteristics have many shared characteristics, but some minor differences. 
5: Extremely similar: The laughter sounds nearly identical.

\begin{table*}[t]
  \caption{
  Evaluation
  of several models for DiariST-AliMeeting laughter test set with Chinese-to-English speech-to-speech translation setting. The best number with a fully automatic system is marked with \textbf{a bold font}. S2ST: Speech-to-Speech Translation, S2TT: Speech-to-Text Translation}
  \label{tab:framework_comparison}
 \vspace{-3mm}
  \centering
{ \footnotesize
\begin{tabular}{@{}cccccccc@{}}
    \toprule 
Speech translation & TTS && ASR-BLEU$\uparrow$ & Speaker SIM-o$\uparrow$& AutoPCP$\uparrow$ & Laughter timing$\uparrow$ & Laughter SIM$\uparrow$\\ 
 \midrule
 \multicolumn{2}{c}{Seamless Expressive (S2ST) \cite{barrault2023seamless}} && {\bf 15.4} & 0.210 & 2.31 & -0.026 & 0.489 \\  
 
 \hdashline[1pt/2pt]\hdashline[0pt/1pt]  
 
                    & Our baseline TTS && {\bf 15.4} & 0.379 & 2.24 & -0.010 & 0.323 \\
Seamless Expressive (S2TT) \cite{barrault2023seamless} & ELaTE (prob.) && 15.0  & 0.383  & 3.06 & 0.661 & 0.750 \\
                    & ELaTE (emb.) && 15.0  & {\bf 0.387}  & {\bf 3.24}  & {\bf 0.673} & {\bf 0.796}  \\
\midrule
                    & Our baseline TTS &&  92.8 & 0.378 & 2.24 & 0.010 & 0.325  \\
Ground-truth translation       & ELaTE (prob.) && 91.5 & 0.385 & 2.97 & 0.618 & 0.727  \\
                    & ELaTE (emb.) && 88.1 & 0.393 & 3.18 & 0.647 & 0.777\\
                \bottomrule
  \end{tabular}
  }
\end{table*}

\subsection{Results on the LibriSpeech test-clean}
In our initial study, we evaluate the impact of fine-tuning on the neutral speech condition using the LibriSpeech test-clean dataset. We input all-zero vectors as the laughter representation to examine any potential side effects of the additional laughter control. 
In this experiment, we mixed the pre-training data
and laughter-conditioned fine-tuning data in a 50:50 ratio during
the fine-tuning stage.
In the evaluation, we generated speech with three different random seeds and took the average of the scores.
In the inference, we applied classifier-free guidance with a guidance strength of 1.0,
and the number of function
evaluations was set to 32.

As indicated in Table~\ref{tab:test-clean}, the fine-tuning process does not affect the WER and offers a slight improvement on the Speaker SIM-o. 
It is crucial to note that the audio quality does not degrade with the inclusion of laughter conditioning. 
In subsequent experiments, we use the fine-tuned TTS system 
without laughter conditioning as our baseline model to ensure a fair comparison.

\begin{figure}[t]
  \centering
  \includegraphics[width=1.0\linewidth]{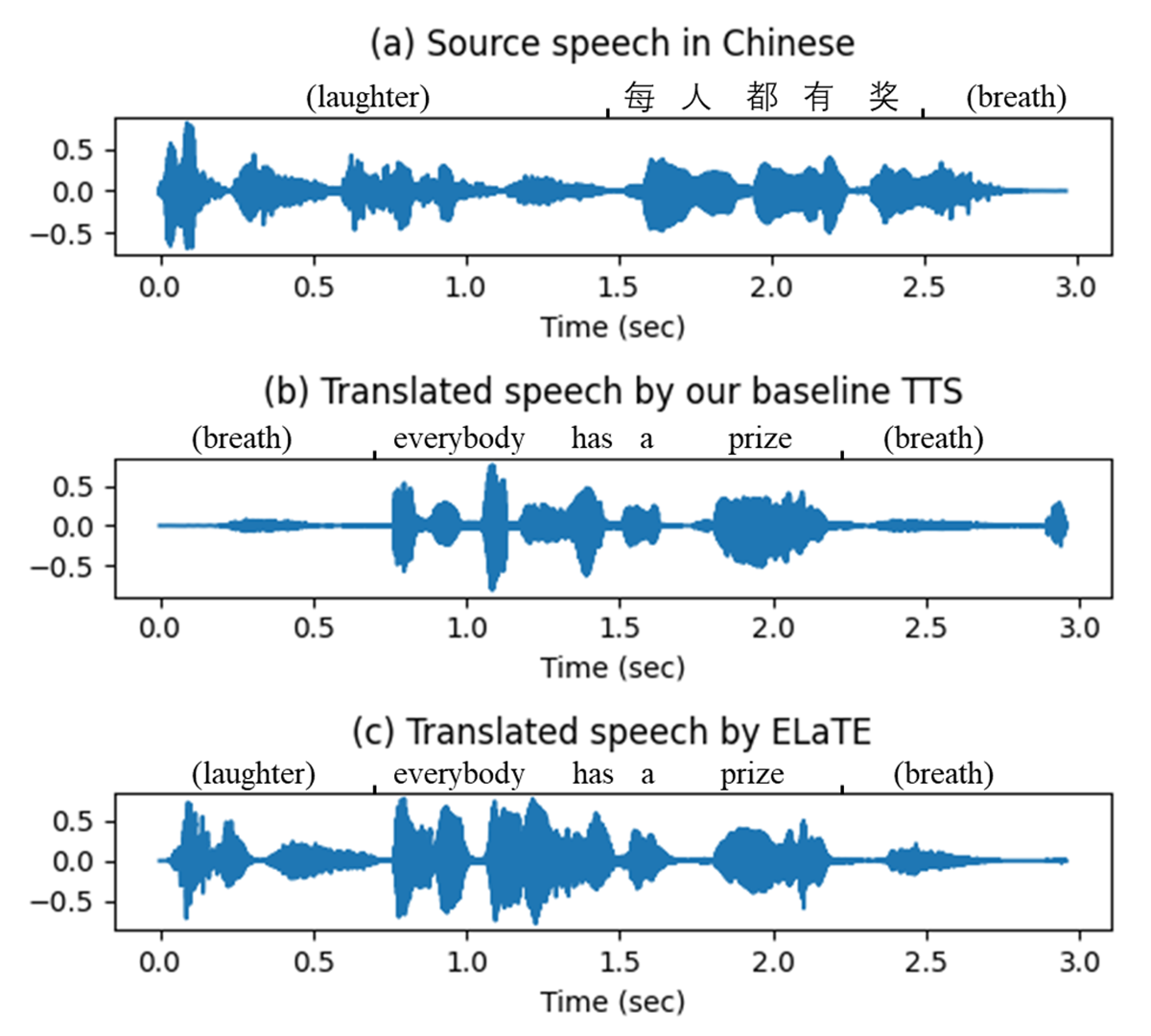}
      \caption{An example of (a) source Chinese audio, (b) translated speech with our baseline TTS, and (c) translated speech with ELaTE.}
  \label{fig:example}
\end{figure}

\begin{table*}[t]
\ra{1.0}
  \caption{Impact of fine-tuning configurations. The best number is marked with \textbf{a bold font}.}
  \label{tab:data}
 \vspace{-3mm}
  \centering
{ \footnotesize
 \tabcolsep = 1.0mm
\begin{tabular}{@{}ccccccccccccc@{}}
    \toprule 
\multicolumn{4}{c}{Fine-tuning configuration} && \multicolumn{2}{c}{LibriSpeech test-clean} && \multicolumn{5}{c}{DiariST-AliMeeting laughter test} \\ \cmidrule{1-4} \cmidrule{6-7} \cmidrule{9-13}
Data & Size (hr) & Ratio & Laughter && WER$\downarrow$& Speaker SIM-o$\uparrow$ && ASR-BLUE$\uparrow$ & SIM$\uparrow$& AutoPCP$\uparrow$ & Laughter timing$\uparrow$ & Laughter SIM$\uparrow$ \\
\midrule
-             & -   & -      & - && {\bf 2.2} & 0.653 &&  91.5 & 0.325 & 2.28 & 0.057  & 0.380 \\
AMI+SW+Fisher & 459.8 & 50\% & - && {\bf 2.2} & 0.658 && {\bf 92.8}  & 0.378 & 2.24 & 0.010  & 0.325 \\
 \hdashline[1pt/2pt]\hdashline[0pt/1pt] 
AMI           & 7.7   & 50\% & $\checkmark$ && {\bf 2.2} & 0.638  && 90.4 &  0.337 & 2.98 & 0.562 & 0.680 \\
AMI+SW        & 41.6  & 50\% & $\checkmark$&& {\bf 2.2} & 0.642  && 91.3 & 0.353 & 3.02  & 0.649 & 0.733 \\
AMI+SW+Fisher & 459.8 & 50\% & $\checkmark$&&  {\bf 2.2} & {\bf 0.662} && 88.1 & {\bf 0.393} & {\bf 3.18} & 0.647  & 0.777   \\
 \hdashline[1pt/2pt]\hdashline[0pt/1pt] 
AMI+SW+Fisher & 459.8 & 25\% & $\checkmark$&& {\bf 2.2} & 0.661 && 90.1 & 0.381 & 3.17 & 0.616 & 0.752 \\
AMI+SW+Fisher & 459.8 & 50\% & $\checkmark$&& {\bf 2.2} & {\bf 0.662} && 88.1 & {\bf 0.393} & {\bf 3.18} & 0.647 & 0.777 \\
AMI+SW+Fisher & 459.8 & 75\% & $\checkmark$&& {\bf 2.2} & 0.659 && 90.2 & 0.384 & {\bf 3.18} & 0.660 & 0.789 \\
AMI+SW+Fisher & 459.8 & 100\% & $\checkmark$&& 2.9 & 0.467 && 86.1 & 0.330 & 3.14 & {\bf 0.704} & {\bf 0.814}  \\
\bottomrule
  \end{tabular}
  }
\end{table*}

\begin{figure}[t]
  \centering
  \includegraphics[width=1.0\linewidth]{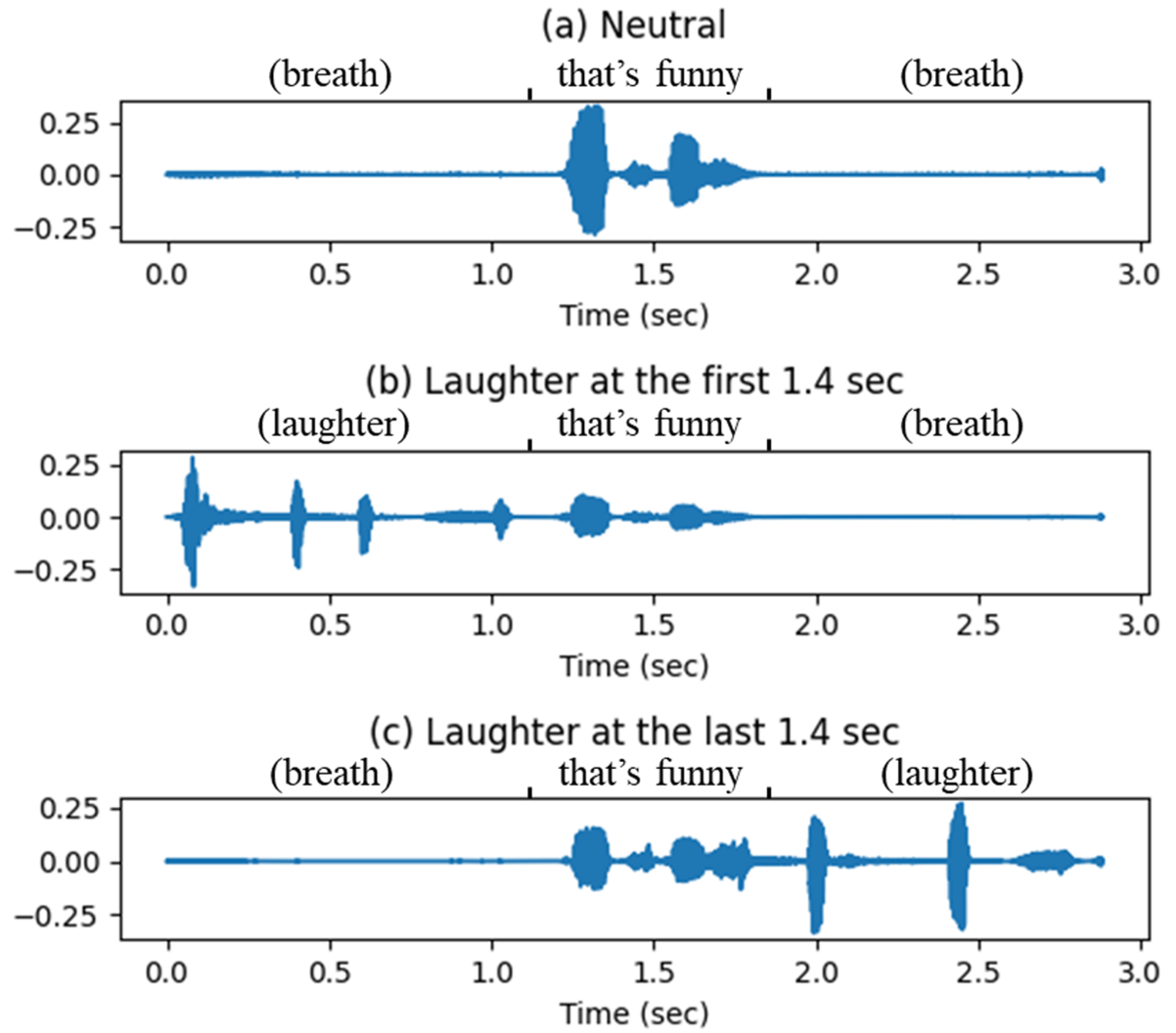}
      \caption{Generated speech by ELaTE with speaker probability where the laughter probability is set to be (a) 0 for all frames, (b) 1 for the first 1.4 seconds and 0 for the rest, and (c) 0 for the first 1.4 seconds and 1 for the rest. The speaker prompt was taken from the speaker 1089 of LibriSpeech test-clean, and the text prompt was set to ``that’s funny''. In this example, we added 1 second of silence frames at the start and end of the estimated duration.}
  \label{fig:thatsfunny}
\end{figure}

\subsection{Results on the DiariST-AliMeeting laughter test set}
\subsubsection{Objective evaluation}

Table~\ref{tab:framework_comparison} presents the results of
the evaluation on the DiariST-AliMeeting laughter test set. In this comparison, we evaluated several systems, including the end-to-end speech-to-speech translation (S2ST) model named Seamless Expressive~\cite{barrault2023seamless}.\footnote{Our experiment is based on Seamless Expressive, supported by the Seamless Licensing Agreement. Copyright © Meta Platforms, Inc. All Rights Reserved.}
For our models, we generated speech with three different random seeds and took the average of the scores.

Our initial observation was that our baseline TTS model achieved a significantly higher Speaker SIM-o compared to the Seamless Expressive S2ST model, despite the absence of laughter conditioning. Interestingly, our baseline TTS model occasionally generates laughter during silent frames. However, the expressiveness measurements for our baseline model, including AutoPCP and Laughter SIM score, were still lower than those of Seamless Expressive. This discrepancy is likely due to the inability of our model to transfer the laughter present in the source audio.
We also noted that the laughter timing score of Seamless Expressive is relatively low. This outcome is likely due to their model architecture, which utilizes a single expressive feature, rendering it unsuitable for transferring laughter timing.

Upon adding both laughter probability and laughter embeddings, we observed significant improvements in Speaker SIM-o, AutoPCP, laughter timing, and laughter SIM. The model utilizing laughter embeddings provided the best results. 
Fig.~\ref{fig:example} shows an example of the source speech, the translated speech by our baseline TTS, and the translated speech by ELaTE with laughter embeddings. As shown in the figure, ELaTE transfers the shape and timing of laughter at the point of a non-speech frame. We can also observe that the output from ELaTE fluctuates more in the speech frame compared to that from the baseline TTS, which mimics the phenomena of laughter while talking.

Note that, while we observed a slightly worse result from the model with laughter probability, it has its own advantage over the laughter embedding model
in that it can control laughter by the direction of time interval to laugh 
as in Fig.~\ref{fig:main} (a). This ability is showcased in Fig.~\ref{fig:thatsfunny}, where we generate multiple speeches with 
different laughter timing instructions.

Finally, we noticed a slight decrease in ASR-BLEU by ELaTE. We speculate that the heavy laughter in some generated speech made it difficult for ASR to recognize, leading to the degradation of the ASR-BLEU.

\begin{figure}[t]
  \centering
  \includegraphics[width=1.0\linewidth]{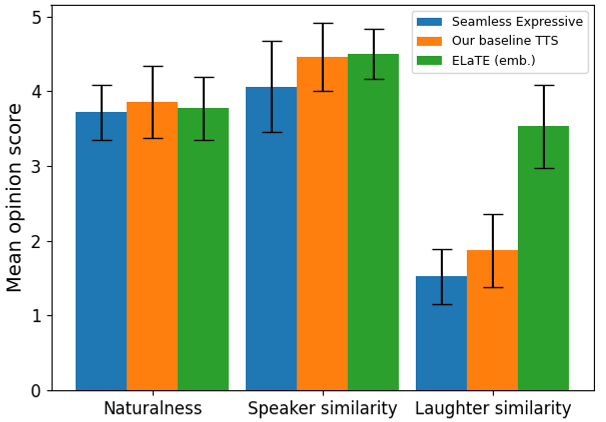}
      \caption{Subjective evaluation results of 30 samples from the DiariST-AliMeeting laughter test set, along with 95\% confidence interval. Naturalness, speaker similarity, and laughter similarity were assessed.}
  \label{fig:mos}
\end{figure}

\subsubsection{Subjective evaluation}

We conducted a subjective evaluation for 30 samples from the DiariST-AliMeeting laughter test set. These 30 samples were selected based on the sentence-level BLEU score, with a preference for higher BLEU scores, while adhering to the constraint that only up to 2 samples could be selected for each speaker.
We asked 12 native English testers 
to rate the metrics described in Section \ref{sec:subjective_metric}.
After collecting subjective scores, we computed z-scores for each score in each sample. A score was considered an outlier if its z-score was greater than 3 or less than -3. If more than 5\% of the scores from a particular tester were outliers, we identified that tester as an outlier. Using this procedure, we excluded one tester for NMOS, one tester for SMOS, and two testers for LMOS. Finally, we calculated the mean and 95\% confidence interval using the crowdMOS tool~\cite{ribeiro2011crowdmos}.

The results are depicted in Fig.~\ref{fig:mos}.
As shown in the figure, ELaTE achieved significantly better laughter similarity compared to both Seamless Expressive and our baseline TTS. Additionally, it is observed that ELaTE maintained naturalness and speaker similarity compared to the baseline TTS, while demonstrating better speaker similarity than Seamless Expressive.

\subsection{Impact of the fine-tuning configuration}

Table \ref{tab:data} illustrates the influence of various training data configurations. In this experiment, we adjusted both the size and the ratio of the laughter-conditioned data during the fine-tuning phase. It's important to note that we proportionally reduced the learning rate schedule in response to a decrease in the size of the training data. This adjustment was necessary due to the significant overfitting observed when the training data size was reduced.

The 3rd and 4th rows of the table reveal a side-effect of fine-tuning when the data size is small. We also observed that a 50:50 mix of laughter-conditioned fine-tuning data and laughter-unconditioned pre-training data yielded the highest model quality.
We believe there is substantial potential for further improvements in the optimization scheme, and exploring these possibilities will be one of our future works.

\section{Conclusions}
In this work, we proposed ELaTE, a zero-shot TTS model capable of generating natural laughing speech from any speaker with precise control over the timing and expression of laughter. More specifically, we fine-tuned a well-trained zero-shot TTS model using a mixture of a small-scale dataset conditioned by laughter signals and a large-scale pre-training dataset where the conditioning signal is set to zero. The resulting model achieves zero-shot TTS capability with precise control of laughter timing and expression without compromising the quality of the base model. Through 
objective and
subjective
evaluations, we demonstrated that ELaTE could generate natural laughing speech with precise controllability, significantly outperforming baseline models. The proposed model represents an important step towards controllable speech generative models. Further research can build on this work by expanding the range of supported speech expressions, such as crying and whispering, and establishing a more sophisticated fine-tuning scheme that works with a smaller amount of fine-tuning data.

\bibliographystyle{IEEEtran}

\bibliography{mybib}

\end{document}